\documentclass[letter,12pt]{scrartcl} 
\usepackage{graphicx}
\usepackage{amsmath}
\usepackage[german,english]{babel}
\usepackage[normalem]{ulem}
\newcommand{\etal}{{\em et al.}}

\newcommand{\mz}[1]{#1}
\begin{document}

\title{\sffamily\bfseries Empirical Features of Congested Traffic States and Their Implications for Traffic Modeling}
\author{Martin Sch\"onhof and Dirk Helbing\\[3mm] 
Institute for Economics and Traffic,\\ 
Dresden University of Technology,\\ Andreas-Schubert Str. 23, 01062 Dresden, Germany}
\maketitle
\begin{abstract}
We investigate characteristic properties of the congested traffic states on a \mz{30} km long
stretch of the German freeway A5 north of Frankfurt/Main. Among the approximately \mz{245}
breakdowns of traffic flow in \mz{165} days, we have identified five different kinds of spatio-temporal
congestion patterns and their combinations. Based on an ``adaptive smoothing method''
for the visualization of detector data, we also discuss particular features of
breakdowns such as the ``boomerang effect'' which is a sign of linearly unstable traffic flow.
Controversial issues such as ``synchronized flow'' or stop-and-go waves are addressed as well. 
Finally, our empirical results are compared with different theoretical concepts and interpretations
of congestion patterns, in particular first- and second-order macroscopic traffic models.  
\end{abstract}

\section{Summary of Previous Models and Empirical Results} \label{Prev}

Understanding traffic dynamics can not only help to identify reasons for bottlenecks. It also contributes to the
development of modern driver and traffic assistance systems aiming at the improvement of 
safety, comfort, and capacity. Progress has been made by empirical studies and theoretical
modelling approaches. Apart from traffic scientists, mathematicians and physicists have
also recently contributed to these fields. Because
of the numerous publications, our introductory overview can only be selective, so that we 
refer the reader to some comprehensive reviews (e.g., Gerlough and Huber, 1975; 
Vumbaco, 1981; Leutzbach, 1988; May, 1990; Brilon {\em et al.}, 1993; Transportation Research Board, 1996;
Gartner {\em et al.}, 1997; Helbing, 1997a; Daganzo, 1997a; Bovy, 1998; Hall, 1999;
Brilon {\em et al.}, 1999; Chowdhury {\em et al.}, 2000b, with a focus
on cellular automata; Helbing, 2001a, containing 800 references; Nagatani, 2002).

\subsection{Modeling Approaches}

The main modeling approaches can be classified as follows: {\em Car-following models}
focus on the non-linear interaction and dynamics of single vehicles. They specify their
acceleration mostly as a function of the distance to the vehicle ahead, the own and relative
velocity (e.g., Reuschel, 1950a, b; Gazis {\em et al.}, 1959, 1961; May and Keller, 1967; Gipps, 1981;
Gibson, 1981; Bando {\em et al.}, 1994, 1995a; Krau{\ss}, 1998; Treiber {\em et al.}, 2000;
Brackstone and McDonald, 2000). {\em Submicroscopic models} take into account even details such as
perception thresholds, changing of gears, acceleration characteristics of specific car types,
reactions to brake lights and winkers (Wiedemann, 1974;
Fellendorf, 1996; Ludmann {\em et al.}, 1997). In favour of numerical efficiency,
{\em cellular automata} describe the dynamics of vehicles in a coarse-grained way 
by discretizing space and time (Cremer and Ludwig, 1986; Biham {\em et al.}, 1992; 
Nagel and Schreckenberg, 1992; Chowdhury {\em et al.}, 2000b). 
{\em Gas-kinetic models} agglomerate over many 
vehicles and formulate a partial differential equation for the spatio-temporal evolution of the vehicle density
and the velocity distribution. While Boltzmann-like approaches (Prigogine and Andrews, 1960;
Prigogine and Herman, 1971; Paveri-Fontana, 1975; Phillips, 1977, 1979a, b; Nelson, 1995; Helbing, 1995c)
are mainly suitable for small densities, Enskog-like approaches (Helbing, 1995d, 1996b, 2001a;
Wagner, 1997a; Klar and Wegener, 1997, 1999a, b; Helbing and Treiber, 1998a; Shvetsov and Helbing, 1999) take into account corrections
due to finite space requirements of vehicles. The main application of gas-kinetic models is the
theoretical derivation of {\em macroscopic traffic equations} for the vehicle density and average velocity.
It is common to distinguish two classes of macroscopic models:
{\em First-order models} such as the Lighthill-Whitham-Richard model
(Lighthill and Whitham, 1955; Richards, 1956) or the Burgers equation (Burgers, 1974) are based on a partial
differential equation for the density and a velocity-density relation or a fundamental diagram 
(flow-density relation) only. {\em Second-order models} contain an additional partial differential equation for
the average velocity and take into account the finite relaxation time to adapt the velocity to changing traffic
conditions (Payne, 1971, 1979; Whitham, 1974; Cremer, 1979; Papageorgiou, 1983; K\"uhne, 1984a; Smulders, 1986;
Kerner and Konh\"auser, 1993; Helbing, 1995b; Helbing and Treiber, 1998a; Lee {\em et. al}, 1998). 
If identical driver-vehicle units are assumed, macroscopic traffic models can be also directly derived
from microscopic car-following models (Payne,1971, 1979a; Nelson, 2000; 
Helbing {\em et al.}, 2002), so that the approximations required in gas-kinetic
derivations can be avoided. This is particularly interesting for a simultaneous 
micro-macro-simulation (Helbing {\em et al.}, 2002), which can be performed on-line based on 
empirical boundary conditions. Finally, {\em mesoscopic} or {\em hybrid traffic models} describe the dynamics of
single vehicles in response to aggregate quantities such as the density (Wiedemann and Schwerdtfeger, 1987;
Schwerdtfeger, 1987; Kates {\em et al.}, 1998), while
{\em queueing models} restrict to the temporal change of numbers of vehicles on larger street sections
as a function of entering and leaving flows (Newell, 1982; Kerner, 2001), 
which is sufficient to determine the travel times (Helbing, 2003).
\par
Each of the above mentioned modelling approaches has its own strengths and areas of applications.
While cellular and macroscopic traffic models are presently in use for the identification and forecast
of the traffic state between measurement cross sections (Nagel, 1996b, 1998;
Esser and Schreckenberg, 1997; Esser {\em et al.}, 1999;
Kaumann {\em et al.}, 2000), microscopic car-following and submicroscopic
models are applied to the development of driver and traffic assistance systems 
(Treiber and Helbing, 2001). In any case,
for most practical applications it is not sufficient to reproduce the relation between flow and density. 
It is equally important to reproduce the various different traffic states formed on freeways. 
This will be the main focus of this paper. 

\subsection{Flow-Density Relation and Wide Data Scattering}

Here, we will shortly summarize important facts required for the discussion in this paper:
The empirical flow-density relation $Q(\rho)$ depends on the location of the measurement section.
Upstream of a bottleneck, it is discontinous and looks
comparable to a mirror image of the Greek letter lambda. The two
branches of this reverse lambda are used to define free low-density and congested
high-density traffic (see, e.g., Koshi \etal, 1983;
Hall \etal, 1986; Neubert \etal, 1999b; Kerner, 2000a; cf. also Edie and Foote, 1958). 
At low vehicle densities, the relationship between flow and density is more or less one-dimensional.
The slope $Q'(\rho)$ at vanishing density $\rho = 0$~veh./km
corresponds to the maximum average speed $V_0$
on the respective road section, but it decreases with increasing density $\rho$ (due to mutual obstructions
during attempted overtaking maneuvers). 
\par
After the flow-density relation $Q(\rho) = \rho V(\rho)$
reaches a maximum at some critical density $\rho = \rho_{\rm cr}$
due to a decrease in the average vehicle speeds $V(\rho)$, 
there is a 
capacity drop (Edie, 1961; Treiterer and Myers, 1974; Ceder and May, 1976;
Payne, 1984), which is related to an increase of
the average (brutto and netto) time gaps (Banks, 1999; Neubert {\em et al.}, 1999; Tilch and Helbing, 2000;
Nishinari {\em et al.}, 2003). This has inspired Hall {\em et al.} to
relate traffic dynamics with catastrophe theory (Hall, 1987; Dillon and Hall, 1987;
Persaud and Hall, 1989; Gilchrist and Hall, 1989).
The flow-density data at higher densities are widely 
scattered in a two-dimensional area (see, e.g., Koshi \etal, 1983;
Hall \etal, 1986; Leutzbach, 1988; K\"uhne, 1991b) 
and erratically varying (Kerner and Rehborn, 1996b) around the so-called jam line 
(Kerner and Konh\"auser, 1994). It can be represented in the form
\begin{equation}
 J(\rho) = \frac{1}{\overline{T}} \left( 1 - \frac{\rho}{\rho_{\rm jam}} \right) \, ,
\label{jamline}
\end{equation}
where the parameter $\overline{T}$ can be interpreted as the average netto time gap of vehicles
leaving congested regions and $\rho_{\rm jam}$
as the density inside of (standing) traffic jams. 
The slope $c = - 1/(\rho_{\rm jam} \overline{T})$ determines the
propagation velocity of perturbations in congested traffic (Kerner, 1998a). Moreover, recent empirical 
investigations of single-vehicle data (see Fig.~(\ref{Nishinari})) have shown that
changes of congested flow $Q$ with time $t$ are very well approximated by 
\begin{equation}
 Q(t) = \frac{1}{T(t)} \left( 1 - \frac{\rho(t)}{\rho_{\rm jam}(t)} \right) \, ,
\label{Qt} 
\end{equation}
and the scattering of ``synchronized'' congested traffic
can be interpreted mainly as an effect of the large variation in the netto time gaps $T$ (Nishinari {\em et al.}, 2003), 
as suggested by Banks (1999). 
\par
\begin{figure}[hptb]
\begin{center}
\end{center}
\caption[]{The two-dimensional scattering of empirical
flow-density data in synchronized traffic flow of high density $\rho \ge 45$ veh/km/lane,
see (a), is well reproduced by the jam relation (\ref{jamline}), when not
only the variation of the density $\rho$, but also the empirically measured variation of
the average time gap $T$ and the maximum density $\rho_{\rm max}$ is taken into account as in
Eq.~(\ref{Qt}), see (b) (from Nishinari {\em et al.}, 2003). 
The pure density-dependence $J(\rho)$ (thick black line) is linear and 
cannot explain a two-dimensional scattering. However, variations of the
average time gap $T$ change its slope $-1/(\rho_{\rm max}T)$ (see arrows), 
and about 95\% of the data are located 
between the thin lines $J(\rho,\overline{T}\pm 2\Delta T,1/l) = (1-\rho l)/(\overline{T}\pm 2 \Delta T)$,
where $l = 3.6$~m is the average vehicle length, $\overline{T}=2.25$~s the average time gap, and
$\Delta T=0.29$~s the standard deviation of $T$.  The predicted form of this area is club-shaped, as
demanded by Kerner (1998b).
\label{Nishinari}}
\end{figure}
In an intermediate density range with $\rho_{\rm c1} < \rho < \rho_{\rm c2}$, 
both free and congested traffic may be found at different times, which points to
hysteresis (Treiterer and Myers, 1974). The traffic states corresponding to the tip of the inverse lambda-shaped
flow-density relation can last for many minutes (see, e.g., Cassidy and Bertini, 1999), but they are not stable,
since it is only a matter of time until a transition to the lower, congested branch of the lambda takes place
(Persaud \etal, 1998; see also Elefteriadou \etal, 1995). The probability $P(\rho)$ of a transition from free to
congested traffic during a given time period is 0 for $\rho = \rho_{\rm c1}$ and becomes
1 at $\rho = \rho_{\rm c2} = \rho_{\rm cr}$. This indicates a metastability of traffic 
flow in the intermediate density range between $\rho_{\rm c1}$ and $\rho_{\rm c2}$
(Kerner and Rehborn, 1996b, 1997; Kerner, 1998b, 1999a, b, c, 2000a, b). 
The transition probability increases with growing density
(Persaud \etal, 1998).

\subsection{Stop-and-Go Waves}

It is possible to distinguish several forms of congested traffic: Stop-and-go waves (start-stop waves) have been
empirically studied by a lot of authors, including Edie and Foote (1958), Mika {\em et al.} (1969), and 
Koshi {\em et al.} (1983). The latter have found that the parts of the velocity-profile
which belong to the fluent stages of stop-and-go waves
do not significantly depend on the flow (regarding their height and length), 
while their oscillation frequency does. 
Correspondingly, there is no {\em characteristic} frequency of stop-and-go traffic. 
The average duration of one wave period is normally between 4 and 20
minutes for wide traffic jams
(see, e.g., Mika \etal, 1969; K\"uhne, 1987; Helbing, 1997a, c, e), 
and the average wave length between
2.5 and 5~km (see, e.g., Kerner, 1998a). Stop-and-go waves
propagate against the direction of the vehicle flow
(Edie and Foote, 1958; Mika \etal, 1969) 
with a velocity $c=-15\pm 5$ km/h (see, e.g.,  Mika \etal, 1969; Kerner and Rehborn, 1996a; 
Kerner, 2000a, b; Cassidy and Mauch, 2001)
and without spreading (Cassidy and Windover, 1995; Windower, 1998; Mu\~noz and Daganzo, 1999;
see also Kerner and Rehborn, 1996a, and the flow and speed data reported by Foster, 1962;
Cassidy and Bertini, 1999). Moreover, wide
moving jams, i.e. moving jams whose width in longitudinal direction is considerably higher than the 
width of the jam fronts, are characterized by stable wave
profiles and by kind of ``universal'' parameters (Kerner and Rehborn, 1996a). Apart from
the propagation velocity $c$, these include 
the density $\rho_{\rm jam}$ inside of jams,
the average velocity and flow inside of traffic jams,
both of which are approximately zero, and
the outflow $Q_{\rm out}$ from jams. Therefore, fully developed traffic
jams can move in parallel over long time periods and road
sections  through free traffic or ``synchronized'' 
congested flow (Kerner, 1998b). The propagation speed of wide jams is not even influenced by
ramps or intersections (Kerner and Rehborn, 1996a; Kerner, 2000a, b).
However, the concrete values of the characteristic parameters slightly depend
on the accepted safe time clearances, the average vehicle length,
truck fraction, and weather conditions (Kerner and Rehborn, 1998a).

\subsection{Queued or ``Synchronized'' Traffic and ``Pinch Effect''} \label{Queued}

The most common form of congestion is not localized like a
wide moving jam, but spatially extended, and it often persists over several hours. In contrast
to stop-and-go waves, the flow and velocity stay finite. Nevertheless, the speed is significantly
reduced, and there is still some capacity
drop (Banks, 1991; Kerner and Rehborn, 1996b, 1998b;
Persaud \etal, 1998; Westland, 1998; Cassidy and Bertini, 1999;
see also May, 1964; Persaud, 1986; Banks, 1989, 1990; Agyemang-Duah and Hall, 1991; Daganzo, 1996).
As the downstream front of this form of congestion is fixed
(e.g. at the location of an on-ramp), it is natural to interpret it as a queuing effect at bottlenecks, when the outflow from 
a road section is exceeded by the inflow, e.g. during rush hours or due to construction sites or accidents.
While Kerner and Rehborn (1996b) call these extended forms of congested traffic
{\em ``synchronized flow''} (because of the synchronization of the velocities 
among lanes), Daganzo (2002b) speaks of 1-pipe flow.
Note, however, that a synchronization of the velocity-profiles in neighboring lanes 
is found for {\em all} congested traffic states including stop-and-go waves or wide moving jams
(Helbing, 1997a). This synchronization
is caused by lane changing maneuvers which equilibrate speed differences among neighboring lanes, when drivers
feel obstructed by congested traffic (Shvetsov and Helbing, 1999; see also the Java simulation 
applet of two-lane traffic on a ring road at {\tt http:/$\!$/www.mtreiber.de}). 
\par
Kerner and Rehborn (1996b) distinguish three different kinds of ``synchronized'' congested traffic:
\begin{itemize}
\item[(i)] {\em stationary and homogeneous states} where both the average
speed and the flow rate are stationary during a relatively long time
interval (see, e.g., also Hall and Agyemang-Duah, 1991; Persaud {\em et al.},
1998; Westland, 1998), 
\item[(ii)] {\em ``homogeneous-in-speed states''} (see also Kerner, 1998b; Lee \etal, 2000)
reminding of recovering traffic (Treiber and Helbing, 1999; Helbing, 2001a),  
which are mainly found downstream of bottlenecks
and are characterized by the fact that only the average vehicle speed is stationary, and
\item[(iii)] {\em non-stationary and non-homogeneous states}
(see also Kerner, 1998b; Cassidy and Bertini, 1999; Treiber {\em et al.}, 2000).
\end{itemize}
\par
The mechanism for the formation of wide moving jams or stop-and-go waves is still
controversial. Treiterer's (1966, 1974) evaluations of vehicle trajectories based on aerial photographs
suggested the existence of {\em ``phantom traffic jams''}, i.e. the
spontaneous formation of traffic jams with no obvious reason such as an accident or a bottleneck.
However, the spontaneous appearance of stop-and-go traffic has recently been questioned.
According to Daganzo (2002a), the breakdown of free traffic 
``can be traced back to a lane change in front of a
highly compressed set of cars'', which shows that there is actually a
reason for jam formation, although its origin can be a rather small disturbance.
Daganzo (2002b) suggests in accordance with Cassidy and Mauch (2001) that 
small oscillations may grow in amplitude due to a pumping effect at ramps. 
In his empirical investigations, Kerner (1998a) finds that jams can be born from extended
congested traffic, which is based on the ``pinch effect'': Upstream of a section with
homogeneous congested traffic close to a bottleneck, there is a
so-called {\em ``pinch region''} characterized by the spontaneous 
birth of small narrow density clusters, which are growing while they travel further
upstream. Wide moving jams are eventually formed by the merging or disappearance of
narrow jams. Once formed, wide jams seem to suppress the occurence of new
narrow jams in between. Similar findings were reported by Koshi {\em et al.} (1983), 
who observed that ``ripples of speed grow larger
in terms of both height and length of the waves as they propagate upstream''.
Note that, instead of forming wide jams,  narrow jams may coexist 
when their distance is larger than about 2.5~km (Kerner, 1998a; Treiber \etal, 2000).

\subsection{Organisation of this Paper}

The mechanism of jam formation will be one of the focusses of this paper. Section~\ref{Data}
will describe our measurement section and its topology, the traffic data, and how we evaluate
them.  In particular, we will propose and use an adaptive smoothing algorithm {as an alternative
to the use of cumulative plots} (Newell, 1982; Cassidy and Windover, 1995; Coifman, 2002; 
Mu\~{n}oz and Daganzo, 2002a; Bertini {\em et al.}, 2003).
It allows us to obtain an intuitive picture of the traffic state on the entire freeway section.
In Section~\ref{States}, we will discuss representative examples of the traffic states and phenomena
that we have identified in our data. Altogether, we will classify free traffic,
five different congested traffic states (including the above mentioned ones), 
and combinations of them. In particular, we will present empirical support for the existence of
growing perturbations, which trigger the breakdown of traffic flow.  According to 
Sec.~\ref{Comparison}, this feature is not
compatible with first-order macroscopic traffic models such as the Lighthill-Whitham-Richards
model (Lighthill and Whitham, 1995; Richards, 1956), while it is reproduced by second-order models. We will also
present a theory (and a phase diagram) of congested traffic states, which is compatible with our empirical findings
and allows one to classify the rich variety of existing traffic models. Finally, Sec.~\ref{Discussion}
summarizes our results and discusses further research directions, e.g. whether it is possible to
reach a synthesis of first- and second-order models.

\section{Investigation Site and Data Evaluation} \label{Data}

During the last year, we have collected data from various freeways around the world in order
to study spatio-temporal phenomena in traffic, to check predictions of traffic models, and to
calibrate their parameters. Most of the data from Germany, The Netherlands, Japan, and the US
are aggregated data from measurement cross sections such as one-minute data for the average
velocity and the vehicle flow, but we have also investigated single vehicle data, data from 
car-following experiments, and floating-car data. Here, we will focus on one-minute
velocity data from a freeway section which frequently suffers from serious congestion and
has also been investigated by Kerner (Kerner and Rehborn, 1996a, b, 1997; Kerner 1998a, b, 1999a, c, 2000a, b, c, 2002). 
Measurements were available for all freeway lanes and most ramps,
but apart from the investigation of exceptional situations such as accidents,
we have arithmetically averaged the speeds over all freeway lanes. 
A sketch of the measurement site is shown in Fig.~\ref{site}. 
It shows an approximately 30 kilometer long freeway section of the German autobahn A5
near Frankfurt/Main, which has mostly 3 lanes into each direction, three intersections with other
freeways, and one junction. At the intersections, there are additional merging and diverging
lanes, some of which are more than 1 kilometer long. Between the intersections ``Frankfurt North-West''
and ``Bad Homburg'', but also between the latter section and the junction ``Friedberg'', there are two
approximately 10 kilometer long three-lane freeway sections without disturbances by on- or off-ramps.
\mz{However, the freeway crosses a valley at kilometer 478 to 480 with gradients of ca. 2-3\%, 
which causes an additional flow-conserving bottleneck in 
direction North. Moreover, there is a relatively steep hill (``K\"opperner Berg'') between kilometers 471 and 472.5 
with gradients up to 5\%, which often produces congestion patterns in direction South.
The main bottlenecks on the considered freeway stretch and their
reasons are listed in Table~\ref{tab_bottleneck_positions}, which also contains the estimated bottleneck strengths.}
\begin{table}[htbp]
\begin{center}
\small
\begin{tabular}{|c|c|c|c|}
\hline
\multicolumn{1}{|c|}{\rule[-4mm]{0mm}{10mm}Position}             & Direction & Reason of Bottleneck & Estimated Strength \\
\hline
{\rule[-4mm]{0mm}{10mm}                   km 471.1}            & South     & Junction Friedberg   &  550 veh./h/lane (max.)        \\
{\rule[-4mm]{0mm}{10mm}                   km 481.3}            & South     & Intersection Bad Homburg  &  320 veh./h/lane         \\ 
{\rule[-4mm]{0mm}{10mm}                   km 488.5}            & North     & Intersection Frankfurt North-West    &  330 veh./h/lane \\ 
{\rule[-4mm]{0mm}{10mm}                   km 478.0}            & North     & Gradient & 110 veh./h/lane \\
\hline
\end{tabular}
\end{center}
\caption{\label{tab_bottleneck_positions}Main bottlenecks on the investigated section of the German
freeway A5 north of Frankfurt.} 
\end{table}

\par
\begin{figure}[htbp]
\begin{center}
\end{center}
\caption{Sketch of the investigated freeway section, showing the two directions of the German three-lane
freeway A5 near Frankfurt/Main. \mz{Each measurement cross section of the freeway is marked by a vertikal line. It is named
with the initial of the driving direction (``N'' for north or ``S'' for south), followed by a number increasing in 
driving direction. The geographical position of each detector is given in kilometers according to the official notation of the
responsible road authorities.}
\label{site}}
\end{figure}
We have investigated the data for both driving directions of the above freeway for all the \mz{165} days
between \mz{04/01/2001 and 09/30/2001}, which allows us to identify the typical features of traffic 
flow and to make statistical analyses. On these days (distinguishing both directions),
we have identified more than \mz{240} breakdowns of traffic flow 
mostly during the morning rush hour between about \mz{7 am and 9 am}  and during the afternoon
rush hour between about \mz{3 pm and 7 pm} (on Fridays between about \mz{1 pm and 7 pm}). Breakdowns were also
induced by holiday traffic or accidents. The latter normally occur outside of the typical bottleneck areas
and are protocolled by the road authorities. During the investigated time period, about 500
accidents have occured. However, most of them had only a minor impact on the traffic flow, as the
cars involved were parked on the emergency lane and did not block any of the freeway lanes.

Our study is based on aggregate double-loop detector data containing, among other information,
the arithmetically averaged vehicle velocities and traffic flows at \mz{30 cross sections
(in direction North) or 31 (in direction South), see Fig. \ref{site}.
We have pre-processed our one-minute data by an error correction which is followed by a smoothing routine.
Apart of the few times in a month, when none of the detectors recorded any 
data for a certain time period due to maintainance work or a breakdown of the computer facilities, 
the data showed very few errors of two types:
(1) Sometimes the data of a cross section just consisted of error bits for a single minute or two subsequent minutes.
In this case (amounting to approx. 1\% of the data), we applied a linear interpolation in time.
If error bits for longer time intervalls than two minutes were found, all values were set to zero. After
visualization of the data, these events were clearly visible, because errors of this type 
occured simultanously for all detectors of a direction.
(2) In other cases, some detector (measuring a single lane at a specific cross section) failed, 
which sometimes lasted for a few weeks. When traffic around this cross section was congested,
we averaged the data in the other lanes. Because of similar velocities in neighboring lanes
during congestion, this procedure leads to realistic velocities.  
If there was free traffic, we interpolated the data of the same lane at the preceeding and the following cross section.  
}
Two representative examples of faulty measurements are displayed in Fig.~\ref{faulty_measurements}.
\begin{figure}[htbp] 
\begin{center}
\end{center}
\caption{\mz{Two Examples of faulty measurements: (a) Between 9:19 and 9:30 there was no data update,
and all detectors delivered permanently the data of 9:19. For the time intervalls 9:30 to 9:33 and 9:37 to 9:40
all detectors delivered error bits which were replaced by zero-values.
This is the reason for the short episode of low velocities over the whole section after 9:30. The zero-values during this intervalls of 4 minutes
cannot be seen clearly because of the applied smoothing (in particular between 9:33 and 9:37 data were available).
(b) Example, where the error bits between 10:09 and 10:39 were replaced by zeros.}
}
\label{faulty_measurements}
\end{figure}
In order to get a spatio-temporal impression of the traffic patterns based on the data of a few
measurement cross sections, we have applied an adaptive smoothing method, which has shown to
deliver realistic results also when the number of measurement cross sections is 
reduced (Treiber and Helbing, 2002). The adaptive
smoothing method uses an exponential filter $\phi(x,t)$ which smoothes over an average time window $\tau$ and
an average spatial interval $\sigma$. In this study, we have always 
chosen \mz{$\tau = 1.2$~min and $\sigma = 0.6$~km.}
The particular feature of the method is the smoothing into the respective propagation direction
of perturbations in traffic flow (i.e. along the ``characteristic lines''). 
In free traffic, perturbations are assumed to propagate forward (downstream) with a speed of
approximately $c_{\rm free} = 80$~kilometers per hour, while in congested traffic, the perturbations
travel upstream with about $c_{\rm cong} =-15$ kilometers per hour (cf. Fig.~\ref{filterSketch}). 
These values have been calibrated in a way that minimizes discontinuities
in the three-dimensional representation of the average velocity along the freeway in the course of time
and agree well with other observations (Kerner and Rehborn, 1996a). The adaptive smoothing method switches
automatically between the free and the congested regime based on a certain criterion, so that there
is no subjective element in this method of data preprocessing (for details see Treiber and Helbing, 2002). The main properties
and advantages of this method are illustrated by Fig.~\ref{vFreevCong}. 
\par
We have applied this method to the representation
of different quantities such as the average velocity $V(x,t)$, its inverse $1/V(x,t)$, the traffic flow $Q(x,t)$, or the vehicle density
determined by $\rho(x,t) = Q(x,t)/V(x,t)$. It turned out that the most intuitive picture of the traffic situation is obtained by
showing the average velocity over space and time, but with high values on the bottom and small values on the 
top. In this way, the data evaluation is based on a quantity which is measured in a relatively easy and reliable way
(in contrast to the vehicle density), and congestion corresponds to hills, similar to the typical representation of the
density. In other words, the higher the value on the vertical axis of the graph, the higher is the resistance to the driver
(the smaller is the velocity). 
\begin{figure}[h!]
\begin{center}
\end{center}
\caption[]{Illustration of the effects of two
linear homogeneous filters with the kernels
$\phi^{\rm free}(x,t)$ and $\phi^{\rm cong}(x,t)$, respectively (from Treiber and Helbing, 2002).
The shaded areas denote the regions considered in the calculation of a data point 
at $(x,t)$. Triangles denote the mainly contributing 
input data sampled in  free  traffic, squares the ones sampled
in congested traffic.\\\\\label{filterSketch}}
\end{figure}
\par\begin{figure}[h!] 
\begin{center}
%
\end{center}
\caption{
\mz{Velocity contour plots of stop-and-go waves observed on 04/05/2001 in direction South.}
Left: Isotropic smoothing, yielding discontinuous patterns. Right: Same data using the adaptive smoothing method,
where travelling waves become clearly visible.
\mz{(For better illustration, just every second detector has been used here, resulting in an effective mean
distance between neighboring detectors of about 2 km. 
The actual mean distance is approximately 1 km, cf. Fig. \ref{site}.) The accident at kilometer 483.1 at 
7:20 is clearly visible in the data.}
\label{vFreevCong}}
\end{figure}
Furthermore, the method smoothes out statistical fluctuations of the measurements
(which are due to the fact that one-minute data average over a small number of vehicles only). In this way, the
main systematic features of traffic patterns become more easily visible. For an investigation of traffic
data from the same freeway stretch with the method of cumulative plots see Bertini {\em et al.} (2003).

\section{Empirical Features of Congested Traffic States} \label{States}

In the following section, we will discuss different kinds of congested spatio-temporal traffic patterns which have
been observed on the investigated section of the German freeway A5. The same patterns have been
identified on sections of other freeways. Differences in the patterns for 
other freeways and other countries will be discussed in Sec.~\ref{Other}.

\subsection{Pinned Localized Cluster (PLC)}

Let us start with a discussion of the particular
kind of spatio-temporal congestion pattern illustrated in Fig.~\ref{PLC}.
This pattern is characterized by a localized breakdown of velocity and a higher density.
Moreover, it has a typical spatial extension. The pattern normally occurs at bottlenecks of the freeway 
during rush-hours and does not move up- or downstream. As it appears to be pinned to a fixed and 
well-defined location, it is named a pinned localized cluster (PLC). In some cases, the cluster may oscillate
in time, which is called an oscillating pinned localized cluster (OPLC), see Fig.~\ref{PLC}b. These oscillating
states may be viewed as spatially confined stop-and-go waves (see Sec.~\ref{Second}).
\par\begin{figure}[htbp]
\begin{center}
\end{center}
\caption{Representative examples of pinned localized clusters.}
\label{PLC}
\end{figure}
Pinned localized clusters may be formed spontaneously (see Fig.~\ref{PLC}a), or they may be caused by
an upstream travelling perturbation which stops at the location of the bottleneck (see Fig.~\ref{PLC}b).
Upstream and downstream of the pinned localized cluster, we have free traffic flow. Pinned localized 
clusters may, therefore, be caused by slower, entering, or lane-changing vehicles
along a bottleneck, e.g. an on-ramp or gradient. When the traffic volume becomes too high, the pinned localized
clusters start to expand in space, which gives rise to other, spatially extended congestion patterns.

\subsection{Homogeneous Congested Traffic (HCT)} \label{hct}

One kind of extended congested traffic is homogeneous congested traffic (HCT). Typically, this pattern
occurs for heavily congested roads, e.g. after serious accidents or during holiday traffic. In Fig.~\ref{HCT}a,
for example, we have  discovered a \mz{complete closing of all three lanes by data analysis,
which has later on been confirmed by the responsible road authorities. 
After a first accident occured at 13:50 at kilometer 477.08, possibly because of an unexpected,
upstream travelling perturbation, 16 other cars were immediately involved in six subsequent accidents.
A quarter of an hour later, two more accidents happened at about the same position. The
congestion in Fig.~\ref{HCT}b was also caused by an accident at 19:15 at kilometer 478.736.} 
\par\begin{figure}[htbp]
\begin{center}
\end{center}
\caption{Representative examples of homogeneous congested traffic.\label{HCT}}
\end{figure}
In homogeneous congested traffic, the (smoothed)
velocity is very low and more or less constant (i.e. homogeneous) over a longer
section of the freeway. The downstream front is located slightly downstream of the upstream end of 
a serious bottleneck, while the downstream end moves upstream, which gives rise to a spatially extended
congestion pattern growing in time. The velocity $C$ of the upstream shock front 
appears to be consistent with the Lighthill-Whitham theory, i.e.
\begin{equation}
 C(t) = \frac{Q_{\rm cong}(t) - Q_{\rm up}(t)}{\rho_{\rm cong}(t) - \rho_{\rm up}(t)} \, ,
\label{shock}
\end{equation}
where $Q_{\rm cong}(t)$ and $\rho_{\rm cong}(t)$ are the flow and density, respectively, 
inside of the congested area, while $Q_{\rm up}$ and $\rho_{\rm up}$ are the (free) flow and density 
in the uncongested area immediately upstream. Downstream of homogeneous congested traffic, 
one usually finds free traffic. Once the bottleneck is removed (e.g. the accident and the lanes blocked by it
are cleared), the downstream front of congested traffic moves upstream with a speed $c$ which approximately
agrees with  $c_{\rm cong} = -15$~km/h. (A small pinned localized cluster of reduced density may, however,
remain at the location of the bottleneck, possibly because of continuing efforts on the emergency lane).
The spatial extension of homogeneous congested traffic shrinks as soon
as $|c| > |C|$ (see Fig.~\ref{HCT}b). The time of disappearance can, in principle be calculated, when the clearing time
and the time-dependent upstream flow $Q_{\rm up}(t)$ are known. The upstream density $\rho_{\rm up}(t)$
may be calculated from the free branch of the fundamental diagram $Q(\rho)$ via the formula
$Q_{\rm up}(t) = Q(\rho_{\rm up}(t))$.

\subsection{Oscillating Congested Traffic (OCT)}

Oscillating congested traffic is another kind of extended congestion pattern. It has similar features
as homogeneous congested traffic regarding its development, growth and dissolution mechanism.
However, the congested area shows more or less regular oscillations of the speed with a 
frequency and amplitude staying relatively constant over a certain period of time. The oscillations
propagate upstream with a velocity $c$ which approximately agrees with $c_{\rm cong}
= -15$~km/h.
\par\begin{figure}[htbp]
\begin{center}
\end{center}
\caption{Representative examples of oscillating congested traffic. The congestion in
subfigure (a) is caused by an accident at kilometer 478.325 at 9:50, while
(b) is a result of a possible hindrance on the fast lane between kilometers 486.0 and 486.9.}
\label{OCT}
\end{figure}
Oscillating congested traffic is often surrounded by free traffic flow and
may be triggered by a perturbation, but it can
also arise when the traffic volume exceeds a certain value. Its downstream front is located slightly
downstream of the upstream end of a bottleneck, until the bottleneck capacity $Q_{\rm cap}$ is sufficient
to cope with the on-ramp flow $Q_{\rm rmp}(t)$ plus the upstream flow $Q_{\rm up}(t)$. The upstream
congestion front propagates again with an average speed given by Eq.~(\ref{shock}), and 
oscillating congested traffic dissolves in a similar way as homogeneous one (see Sec.~\ref{hct}). One
may distinguish two cases: 
\begin{itemize}
\item If the average flow $Q_{\rm cong}(t)$ inside of the congested area
plus the ramp flow $Q_{\rm rmp}(t)$ per lane drop below the capacity $Q_{\rm cap}$ of the
activated bottleneck,
the latter can cope with the overall traffic volume (and becomes
inactive). {\em In this case, the downstream congestion 
front starts to move upstream} with a speed $c \approx c_{\rm cong}$, until the upstream front
is reached and the congestion is thereby dissolved (see Fig.~\ref{OCT}a).
\item If the overall traffic volume $Q_{\rm cong}(t) + Q_{\rm rmp}(t)$ stays above the bottleneck capacity
$Q_{\rm cap}$ during the whole congestion period, the upstream congestion front starts to move downstream 
as soon as the time-dependent upstream flow $Q_{\rm up}(t)$ drops below the 
average congested flow $Q_{\rm cong}$ (see Fig.~\ref{OCT}b).
\end{itemize}
 
\subsection{Stop-and-Go Waves (SGW)}

Another form of congested traffic, which has spatially extended and localized features at the same time,
are stop-and-go waves. They are related to oscillating congested traffic, but they have a large characteristic
amplitude, while there is no typical wavelength. Stop-and-go waves (SGW) consist of a sequence
of traffic jams with free traffic in between. The traffic jams are localized (i.e. spatially confined) and propagate
upstream with velocity $c_{\rm cong} \approx -15$~km/h. The spatial and temporal distance among two 
successive traffic jams varies significantly. Stop-and-go waves are sometimes triggered by small
perturbations in the traffic flow or may originate from an area of pinned localized clusters
(see Figs.~\ref{TSG}, \ref{boom}, and \ref{combi}).
The propagation speed of stop-and-go waves does not change when they travel through
pinned localized clusters (see Figs.~\ref{boom}b, \ref{combi}b) 
or spatially extended congested traffic (Kerner 2000a, b, 2002). 
Their propagation typically ends in free traffic. However, they may also end as pinned localized clusters
(cf. Fig.~\ref{PLC}b). 
\begin{figure}[htbp]
\begin{center}
\end{center}
\caption{Representative examples of stop-and-go waves, with an accident in (a) at kilometer 482.8 at 16:20.}
\label{TSG}
\end{figure}

\subsection{Moving Localized Cluster (MLC)}

In the case of a single moving traffic jam (instead of sequence of them), we talk about a moving localized
cluster (MLC). Compared to a pinned localized cluster, the extension of a MLC state is also limited,
but it is propagating with the speed $c_{\rm cong}$ rather than staying at a particular place. A moving
localized cluster is usually born from a perturbation of traffic flow. One can distinguish two different cases:
\begin{itemize}
\item The perturbation is large enough to travel upstream from the very beginning (see Fig.~\ref{MLC}a).
\item The initial perturbation is small, and one observes a boomerang effect (cf. Fig.~\ref{MLC}b and the
next subsection).
\end{itemize}
\begin{figure}[htbp]
\begin{center}
\end{center}
\caption{Representative examples of moving localized clusters. Note that, in the right picture,
a small perturbation triggers a pinned localized cluster while the moving localized cluster passes the bottleneck
at kilometer 480.}
\label{MLC}
\end{figure}

\subsection{The Boomerang Effect}

One typical mechanism for the triggering of the above mentioned congested traffic patterns is the
so-called boomerang effect (Helbing, 2001b, Helbing {\em et al.}, 2003), 
which can be seen in simulation results of traffic models with a
linearly unstable density range (Kerner and Konh\"auser, 1993). 
According to the boomerang effect, small perturbations
in free traffic propagate downstream, but if they grow, they change their propagation speed and direction,
so that they eventually return. If there is a mechanism which, under certain conditions, 
causes a growth of small perturbations, such a behavior is expected: Small perturbations correspond to clusters of
vehicles, which should move into the direction of the vehicles, i.e. downstream. 
On the other hand, once a moving traffic jam has developed, it is supposed to move in upstream direction:
Inside of the traffic jam, vehicles are standing,  at its downstream front, vehicles are leaving,
and at the upstream front, new vehicles are joining the standing vehicle queue. Consequently, when a
moving traffic jam is born from a perturbation of the traffic flow, the propagation direction should change
and eventually turn around. 
\par
In Sec.~\ref{Second}, we will discuss whether the growth of small perturbations
can be explained in a different way, for example based on ramp flows. Here, we just note that, although
a breakdown of traffic flow is not always related to the boomerang effect, it occurs rather frequently
(see Figs.~\ref{PLC}b, \ref{boom}, and \ref{combi}b). However, it is hard to recognize without
a suitable method of data preprocessing. When no smoothing method is applied, 
small perturbations are normally masked by the fluctuations. Moreover, if the propagation direction of
perturbations is not considered by the interpolation procedure, continuously moving patterns are
splitted up into artificial, discontinuously looking structures, and it is hard to make 
sense of these (see Fig.~\ref{vFreevCong}) It is also
hard to identify the boomerang effect without a three-dimensional representation, just on the basis
of discrete detector data. Therefore, the empirical identification of the boomerang effect 
has profited from advanced methods of data preprocessing such as the adaptive smoothing method
discussed in Sec.~\ref{Data}.
It is highly unlikely that this effect is an artefact of the data representation, as it can sometimes also be
seen in classical contour plots without particular preprocessing (Helbing {\em et al.}, 2003) and in
three-dimensional plots without a spatial smoothing (see Fig.~\ref{ampli}). 
\begin{figure}[htbp]
\begin{center}
\end{center}
\caption{Representative examples of the boomerang effect (see the area around kilometer 485 and
Figs.~\ref{PLC}b, \ref{combi}b).}
\label{boom}
\end{figure}

\begin{figure}[htbp]
\begin{center}
\end{center}
\caption{
Illustration of growing perturbations on a homogeneous freeway section without on- and
off-ramps. Growth of a small, downstream travelling perturbation and boomerang effect represented 
(a) by the adaptive smoothing method and
(b) by time-dependent velocity data obtained at the different measurement cross sections (without
any spatial smoothing), as seen from behind. (c), (d) Examples for the growth of
medium-sized perturbations propagating upstream.
}
\label{ampli}
\end{figure}

\subsection{Combined States and the ``Pinch Effect''}

Freeways are not at all spatially homogeneous. Due to the existence of on- and off-ramps,
gradients, curves etc., a freeway can be imagined to be composed of road sections of different
capacities, even if the number of lanes is constant. If the capacity is reduced from one road section
to the next one in downstream direction, we speak of a bottleneck. Upstream of bottlenecks, there
is a danger of queue formation or congestion, if the traffic volume becomes too high, while downstream
of the upstream end of a bottleneck, one mostly observes free traffic, if there is not another bottleneck
downstream. Congestion normally starts to
form upstream of the freeway section, for which the difference between the arriving traffic volume
and the section capacity (bottleneck capacity) is highest. With increasing traffic volume, more and
more bottlenecks are activated. Therefore, we normally have a sequence of different congested
traffic patterns along the freeway. 
\par\begin{figure}[htbp]
\begin{center}
\end{center}
\caption{Combinations of different congested traffic states.}
\label{combi}
\end{figure}
{For example, Figure~\ref{boom}a shows the formation of a pinned
localized cluster upstream of kilometer \mz{471.5} at about 6:30 am. About 30 minutes later, this state
turns into stop-and-go traffic, and a weaker bottleneck at kilometer 474 is sometimes activated as well.
Around 7:50 am, we observe the formation of a moving localized cluster, which is triggered by 
the boomerang effect. The moving localized cluster turns into a pinned localized cluster at
kilometer 482, which later on triggers pinned localized clusters and stop-and-go waves at kilometer 479
(possibly because of an accident at kilometer 479.2 at 8:50) and at kilometer 469.
\par
In Figure~\ref{boom}b, the boomerang effect triggers a moving localized cluster at about 7:00am. Around 7:30am, 
a pinned localized cluster is formed at kilometer 480, which also appears to be triggered by a small perturbation
which moves downstream and hits the moving localized cluster around kilometer 480. The moving localized
cluster continues, while the pinned localized cluster emits several traffic jams, so that we face triggered stop-and-go
waves. Some of the traffic jams propagate through the pinned localized cluster which has formed around 7:30am at
kilometer 470 and also generates stop-and-go waves. After 9:00am, some of the stop-and-go waves and
the pinned localized traffic at kilometer 470 start to disappear, while the pinned localized cluster at kilometer
480 is still alive at 9:40am.
\par
Figure~\ref{combi}a shows stop-and-go waves upstream of kilometer 480. Around 17:20, homogeneous
congested traffic forms at kilometer \mz{488}, and the downstream flow is considerably reduced. This is a
result of an accident at kilometer 487.5 at 17:13. One hour later, the accident has been cleared. This turns
the homogeneous congested traffic into less severe oscillating congested traffic. Moreover, the downstream
flow is back to normal, which causes a queue upstream of the bottleneck at kilometer 480. 
The congestion pattern emerging over there appears to be oscillating congested traffic as well.
\par
Figure~\ref{combi}b illustrates the formation of a moving localized cluster via a
boomerang effect around 7am. While the cluster passes along the bottleneck at kilometer
478, it meets a downstream moving perturbation, which triggers the formation of a
pinned localized cluster. Around 8am, the moving localized cluster arrives at the bottleneck at
kilometer 471.5 where it triggers oscillating congested traffic that persists until 9 o'clock.
Around 8:30 am, the pinned localized cluster at kilometer \mz{482} turns into extended congested
traffic, and the resulting pattern reminds of a pinch effect.}
\par
Figure~\ref{pinch} may be viewed as an illustration of the so-called pinch effect. After 7:00am,
we find something like homogeneous congested traffic or a pinned localized cluster at kilometer 480,
which turns into oscillating congested traffic further upstream. Some of these oscillations disappear
as they travel upstream, while two of them form stop-and-go waves. These enter another area of
congested traffic upstream of kilometer 471.5. Note that around kilometer 478, we do not really see a merging of 
small oscillations in favour of a few remaining moving traffic jams, in contrast to the suggestion by Kerner (1998a). 
The oscillations rather seem to disappear, i.e. they seem to be dissolved in free traffic. Moreover, the
pinch effect was not frequently observed by us. However, both of this may be due to the applied smoothing procedure. 
Structures on a smaller scale than \mz{$\sigma = 0.6$~km} may actually merge, but from our point of view, it is
hard to tell them apart from fluctuations.
\begin{figure}[htbp]
\begin{center}
\end{center}
\caption{Representative example of the ``pinch effect'', i.e. a spatial sequence
of homogeneous congested, oscillating congested, and stop-and-go traffic in upstream direction.}
\label{pinch}
\end{figure}

\subsection{Comparison with Kerner's Observations}

Comparing our empirical results with Kerner's findings for the same freeway section,
we suggest the following identifications:
Homogeneous congested traffic (HCT) seems to be the same as ``sychronized'' traffic flow (ST) of
type (i) (see Sec.~\ref{Queued}), while  oscillating congested traffic (OCT) seems to relate to
synchronized flow of type (iii). Finally, homogeneous-in-speed states remind of
the free branch of the fundamental diagram, but with a reduced free velocity. Therefore, this
synchronized flow of type (ii) bears
features of both, free and congested traffic. This may point to recovering traffic 
downstream of bottlenecks (see Fig.~4 in Treiber and Helbing, 1999, and Fig.~3 in 
Tilch and Helbing, 2000) or to frustrated
drivers, who have decreased their desired speed after having spent a considerable amount of time in
congested traffic (see Treiber and Helbing, 2003).

\subsection{Occurence Frequency of Traffic States and Relevance for Other Sites} \label{Other}

We have also investigated the occurrence frequency of the above mentioned traffic states
on the investigated freeway stretch. The result
is shown in Figs.~\ref{fig:A5_K_states_histogramm} and \ref{fig:A5_F_states_histogramm}. 
\begin{figure}[h]
\hfill
    \begin{minipage}[t]{0.45\linewidth}
      \caption{Absolute frequencies of congested traffic states 
on the German freeway A5 close to Frankfurt in direction North.
\label{fig:A5_K_states_histogramm} 
}
    \end{minipage}\hfill
    \begin{minipage}[t]{0.45\linewidth}
      \caption{Absolute frequencies of congested traffic states in direction South.
\label{fig:A5_F_states_histogramm} 
}
    \end{minipage}\hfill
\end{figure}
According to this, oscillating congested traffic and pinned localized clusters
are very common on the studied freeway stretch. Homogeneous congested traffic is less frequent than
oscillating congested traffic, in accordance
with expectations. If it occured frequently, the freeway capacity would not be appropriately dimensioned to
cope with the traffic volume. We should, however, note that pinned localized clusters cannot always
be clearly distinguished from homogeneous congested traffic, if the upstream extension of the latter is
limited by a section of high capacity or  an off-ramp, where many vehicles (can) leave. This is, for example,
potentially the case around kilometer 481. In some particular cases, it is also difficult to exactly tell apart 
moving localized clusters from stop-and-go waves, or the latter from oscillating congested traffic. Nevertheless,
it is helpful and makes sense to distinguish the five congestion patterns mentioned above. 
\par
We should mention that one can notice a typical dependence of the frequency of congestion on
the day of the week (see Figs.~\ref{fig:A5_K_states_histogramm} and \ref{fig:A5_F_states_histogramm}). 
On Saturdays and Sundays, there is little congestion compared
to working days, as expected. Moreover, in direction North, congestion patterns appeared seldomly on 
Mondays compared to Tuesdays, Wednesdays, or Thursdays (Helbing {\em et al.}, 2003), while in
direction South, the frequency of traffic breakdowns was surprisingly low on Fridays. This is probably
due to commuters spending their weekends in the North and living in Frankfurt during the week.
\par
It should be mentioned that only less than \mz{five} percent of congestion patterns remained unexplained. The
remaining ones were probably caused by accidents, or they were cases of forwardly moving 
``phantom'' bottlenecks (Gazis and Herman, 1992; Mu\~{n}oz and Daganzo, 2002b), 
which were found two times (see Fig.~\ref{phantom}). 
\begin{figure}[htbp]
\begin{center}
\end{center}
\caption{Congestion pinned at moving bottlenecks, the velocity is about 1.6 km/h (left) and quite exactly 2.7 km/h (right).}
\label{phantom}
\end{figure}
Although we could not anymore identify the exact reason for the
forwardly moving congestion patterns,  it is plausible to assume an exceptionally slow heavy goods 
or road work vehicle (for example, cutting grass along the freeway).
In summary, it can be said that 
the great majority of congestion patterns on the studied freeway section was related to bottlenecks, but some of
them were triggered by small perturbations, e.g. via the boomerang effect \mz{(in 18 out of 245 cases)}.
\par
Let us now shortly address the question, whether the above findings are also relevant for other
freeway sites and other countries. Very similar congestion patterns have been found on other German freeways 
(e.g. A3, A8, A9) or on freeways in the Netherlands (e.g. the freeways A2, A9).  However, the
relative frequency of the different congestion patterns is site-dependent. It seems to depend on
the bottleneck strengths and the respective traffic volumes. Moreover, although there are a few reports of
oscillating features of congestion in the United States (Daganzo {\em et al.}, 1999;
Cassidy and Bertini, 1999), it appears that oscillating congested traffic and
stop-and-go waves occur less frequent than, for example, in Germany. This shows that the frequency of
congestion patterns is also a matter of country and driver behavior, 
probably because parameters such as the
average time gap between vehicles, the acceleration behavior of drivers, the lane changing
frequency and traffic regulations matter, as well as speed limits and the velocity variance.
In any case, we are not aware of additional congestion patterns
in the US, which have not been covered by the above classification. In Japan, by the way, oscillating 
congestion patterns (OCT, SGW) seem to exist (Koshi {\em et al.}, 1983). 
\par
We should finally note that the observations also depend to a certain extent on the detection technology. For example,
pinned localized clusters may be overlooked, if detectors are not placed at suitable locations. Moreover, bottlenecks
can have various origins: Apart from classical bottlenecks such as on-ramps, gradients and accidents, 
there are bottlenecks due to off-ramps (caused by frequent lane changes, weaving flows), curves, bad weather,
changes in illumination (tunnel entrances, blinding sun), or circumstances that reduce the attention of the drivers 
to road traffic (beautiful views, accidents on the opposite lanes), etc.

\section{Comparison with Traffic Models} \label{Comparison}

The identification of empirical traffic patterns is not only of importance to increase the comfort, efficiency and
safety of freeway traffic by technological means, e.g. driver assistance systems. It is also helpful to verify
traffic models used to develop and assess these measures, to identify the traffic state between 
measurement sections, and to forecast traffic. We will, therefore, discuss in the following, to what extent the
above observations are consistent with first- and second-order macroscopic traffic models
and their microscopic equivalents (e.g. car-following models).

\subsection{The Lighthill-Whitham-Richard Model (LWR Model)}

Let us begin with the fluiddynamic model by Lighthill and Whitham (1955) and Richard (1956). It
is based on the continuity equation
\begin{equation}
 \frac{\partial \rho(x,t)}{\partial t} + \frac{\partial Q(\rho(x,t))}{\partial x} = \nu_+ (x,t) - \nu_-(x,t)\, ,
\label{cont}
\end{equation}
where $\rho(x,t)$ denotes the vehicle density as a function of location $x$ and time $t$.
For the time being, we will set the source terms $\nu_{\pm}(x,t)$ due
to ramp flows equal to 0 (where the plus sign
corresponds to on-ramps and the minus sign to off-ramps). 
The model is closed by assuming a flow-density relationship $Q(\rho)$, which is 
called the fundamental diagram. The resulting fluid-dynamic traffic
model assumes an instantanous adaptation of the average speed to the density, as it does not contain
an independent partial differential equation for the dynamics of the average velocity. Therefore, it is
called a first-order model.
\par
We can rewrite Eq.~(\ref{cont}) as a non-linear wave equation
\begin{equation}
 \frac{\partial \rho(x,t)}{\partial t} +
 \frac{dQ(\rho)}{d\rho}  \frac{\partial \rho(x,t)}{\partial x} = 0 \, .
\label{wave}
\end{equation}
According to this, the density profile propagates with the speed
\begin{equation}
 c(\rho) = \frac{dQ(\rho)}{d\rho} \, , 
\label{propa}
\end{equation}
i.e. a formal solution of Eq.~(\ref{wave}) is given by
\begin{equation}
 \rho(x,t) = \rho\bigg(x - \int_0^t \!\! dt' \; c\Big(\rho\big(y_x(t'),t'\big)\Big) , 0\bigg) \, ,
\end{equation}
where 
\begin{equation}
 y_x(t') = x - \int_{t'}^t dt^{\prime\prime} \, v(t^{\prime\prime}) 
 \quad \mbox{with} \quad \frac{dy_x(t')}{dt'} = v(t') = c\Big(\rho\big(y_x(t'),t'\big)\Big) 
\end{equation}
is the location with vehicle density $\rho$ as a function of time $t'$.
As this defines an implicit equation for the density $\rho(x,t)$, it is hard to solve explicitly. However,
we can see from it that the initial density profile $\rho(x,0)$ 
changes its shape in the course of time if the propagation
velocity $c(\rho)$ is density-dependent, but it does not change its amplitude. This is the reason
for the evolution of shock fronts, which propagate with the velocity specified in Eq.~(\ref{propa})
(Whitham, 1974).
\par
The first-order model (\ref{cont}) is suitable to describe the propagation of the upstream
shock fronts of spatially extended congestion patterns such as homogeneous or oscillating congested
traffic. It may also describe the propagation of {\em fully developed} traffic jams, both moving localized clusters
and stop-and-go waves (Helbing, 2003). However, it cannot describe the {\em emergence} of these patterns and of
oscillating congested traffic without 
particular assumptions. Such kinds of assumptions have been formulated. For example, 
Cassidy and Mauch (2001) and Daganzo
(2002b) suggest that these patterns are produced by ramp flows. In particular, they
propose a pumping effect at ramps which may explain structures of increasing amplitude.
Such an assumption makes sense for freeways in the Unites States, where on- and off-ramps often have 
very short distances from each other. In Europe, however, there are freeway sections of about 10 kilometer length without
any ramps or changes in the number of lanes. On these sections, according to the above theory,
we should not find patterns of increasing amplitude
(where we define the amplitude as the difference between the minimum and maximum value of the pattern).
This does, by the way, not only apply to the vehicle density $\rho(x,t)$, but also to the average vehicle
velocity $V(x,t) = Q(\rho(x,t)) / \rho(x,t)$. As $V(\rho) = Q(\rho)/\rho$ is
a monotonically falling function of the density $\rho(x,t)$,
minima of the vehicle density correspond to maxima of the average velocity and vice versa. 
\par
The boomerang effect questions the first-order model (\ref{cont}), as it shows that patterns of growing amplitude
exist even on freeway sections without ramps. Figure \ref{ampli} illustrates this in more detail by comparing
the profiles at subsequent measurement cross sections without any spatial interpolation. As a conclusion, 
our empirical results call for an extension of the Lighthill-Whitham theory. This should be able to describe patterns 
with growing amplitudes and to explain 
emergent oscillating structures without having to assume oscillations in the ramp flows
triggering them. Figure \ref{fourier} shows that the oscillation frequency of triggered oscillating congested traffic
or stop-and-go waves is not at all comparable with the frequency of the variations in the ramp flow. 
\begin{figure}[htbp]
\begin{center}
\end{center}
\caption{Oscillating congested traffic on 07/16/2001 at the Junction Friedberg. 
The oscillation frequency of the on-ramp flow at kilometer 470.8 (upper left) is much higher than
the frequency of the flow approximately 2.4 kilometers upstream (upper right). This visual impression is 
confirmed by the corresponding Fourier spectra (lower figures).
The on-ramp flow has a typical frequency larger than 0.3 per minute, while the
upstream flow shows a typical frequency of less than 0.1 per minute. Therefore, the frequency of
oscillating congested traffic is not determined by the variations in the on-ramp flow.}
\label{fourier}
\end{figure}

\subsection{Phase Diagram of Traffic States for Second-Order Models} \label{Second}

Compared to the Lighthill-Whitham-Richard model, second-order macroscopic traffic models
 do not assume an immediate adaptation of the vehicle velocity to a changing traffic situation.
They contain an additional partial differential equation for the spatio-temporal change of the
average velocity $V(x,t)$ of vehicles, which can often be written in the form
\begin{equation}
 \frac{\partial V}{\partial t}  
 + \underbrace{V \frac{\partial V}{\partial x}}_{\rm Transport\ Term}
 = \underbrace{- \frac{1}{\rho} \frac{\partial {P}}
 {\partial x}}_{\rm Pressure\ Term}
 + \underbrace{\frac{1}{\tau} ( V_{\rm e} - V )}_{\rm Relaxation\ Term} \! .
\label{geschwin}
\end{equation} 
Herein, $P$ is called the traffic pressure, $\tau$ the relaxation or adaptation time $\tau$, and $V_{\rm e}$
the ``optimal'' (or dynamic equilibrium) velocity $V_{\rm e}$, which is dependent 
on the local vehicle density $\rho$ and possibly on other variables as well.
Notice that the Lighthill-Whitham model results in the limit $\tau \rightarrow 0$, but Whitham
has proposed a second-order extension himself (1974).
The models by Payne (1971) and Papageorgiou (1983) 
are obtained for ${P}(\rho) = 
[V_0 - V_{\rm e}(\rho)]/(2\tau)$, with the
``free'' or ``desired'' average velocity $V_0 = V_{\rm e}(0)$. For
$d{P}/d\rho = - \rho/[2\tau (\rho + \kappa)] dV_{\rm e}/d\rho$,
one ends up with Cremer's (1979) model. 
In the model of Phillips (1979), there is
${P} = \rho \theta$, where $\theta$ denotes the velocity variance.
The model of K\"uhne (1984, 1987) and of Kerner
and Konh\"auser (1993) results for
${P} = \rho \theta_0 - \eta \partial V/\partial x$, where
$\theta_0$ is a positive constant and $\eta$ a viscosity coefficient.
In comparison with a similar model by Whitham (1974),
the additional contribution $- \eta \partial V/\partial x$ implies a
viscosity term $(\eta/\rho) \,\partial^2 V/\partial x^2$. This is essential
for smoothing shock fronts, which is desireable from an empirical and
numerical point of view. 
\par
Second-order models have been seriously criticized by Daganzo (1995c),
but his criticism has been overcome by improved macroscopic traffic models
(Helbing, 1995a, b, c, 1996, 1997a, 2001a; Aw and Rascle, 2000; Helbing {\em et al.}, 2001;
Aw {\em et al.}, 2002). The non-local, gas-kinetic-based
traffic (GKT) model, for example, is theoretically consistent, numerically
efficient, realistic, and has been successfully  applied for traffic state
identification in large freeway networks. Instead of introducing a smoothing effect via a viscosity term,
it considers the reaction of drivers to the traffic situation ahead of them (i.e. anticipation
effects). This basically causes an additional dependence of the ``optimal velocity''
from the density and average velocity at an advanced location, which makes the
model non-local. It also takes into account effects of space requirements by vehicle lengths
and safe time clearances. The traffic pressure is not only density-dependent, but
proportional to the squared average velocity, which avoids the possibility of negative velocities.
Correlations between the velocities of successive cars can
be easily treated as well. Here, we will not repeat the detailed mathematical form and properties of this model, 
as they have been described in several references (Helbing and Treiber, 1999; Helbing, 2001;
Helbing {\em et al.}, 2001, 2002). 
Moreover, qualitatively the same traffic states and preconditions for their occurence 
have been found for other second-order models such as
the Kerner-Konh\"auser model (Lee {\em et al.}, 1999). We will only summarize the main points shortly
in this paper, as detailed discussions are available elsewhere 
(Helbing and Treiber, 1999; Helbing, 2001; Helbing {\em et al.}, 2001, 2002). 
\par
The gas-kinetic-based traffic model produces free traffic (FT) and five different kinds of congested traffic states,
which are displayed in Fig.~\ref{Trafstates}. 
\begin{figure}[htbp]
\begin{center}
\end{center}
\caption{\mz{Simulation of a freeway with an on-ramp at location $x=5$~km and an initial disturbance travelling
upsteam, using the non-local, gas-kinetic-based traffic (GKT) model
(Helbing {\em et al.}, 1999). Five different kinds of congested traffic states may emerge,
depending on the respective traffic flows on the ramp and on the freeway:
a moving localized cluster (MLC), a pinned localized cluster (PLC), 
stop-and-go waves (SGW), oscillating congested traffic (OCT), and 
homogeneous congested traffic (HCT). During the first minutes of the simulation,
the flows on the freeway and the on-ramp were increased from low values
to their final values.}}

\label{Trafstates}
\end{figure}
The typical preconditions for their occurence can be illustrated by a phase diagram 
(see Fig.~\ref{Phased}). Each area of the phase diagram
represents the parameter combinations, for which certain kinds of traffic state can exist.
It is interesting that the borderlines between different areas  (the so-called phase boundaries)
can be theoretically understood based on the instability diagram and the dynamic
capacity $Q_{\rm out}$. The latter is given by the outflow from congested traffic and has typical
values of $1800 \pm 100$ vehicles per hour and lane (Kerner and Rehborn, 1996a).
\par\begin{figure}[htbp]
\begin{center}
\end{center}
\caption{Theoretical phase diagram of traffic states with numerical estimates of the
boundaries for the bottleneck around kilometer 470 on the German freeway A5 in direction South. 
The different areas indicate, for which combinations of the upstream freeway flow
$Q_{\rm up}$ and the bottleneck strength $\Delta Q$ certain traffic states may exist,
depending on the initial and boundary conditions. Most areas are multistable, i.e. one may
find one out of several possible states. 
}
\label{Phased}
\end{figure}
The non-local, gas-kinetic-based traffic
(GKT) model and some other second-order traffic models predict stable traffic flow,
when the velocity changes little with the density. More specifically, there is stable traffic below
some critical density $\rho_{\rm c1}$ and above some critical density $\rho_{\rm c4}$
(see Fig.~\ref{instabdiag}). 
\par\begin{figure}[htbp]
\begin{center}
\end{center}
\caption{Schematic illustration of velocity $V$ and flow $Q$ as a function of the vehicle density $\rho$. 
Grey regions indicate density ranges of metastable traffic (cf. text).}
\label{instabdiag}
\end{figure}
For medium
traffic densities between the critical densities $\rho_{\rm c2}$ and $\rho_{\rm c3}$, traffic flow
is linearly unstable, i.e. even the smallest perturbation can grow and cause a breakdown
of traffic flow. In the intermediate density ranges $\rho_{\rm c1} \le \rho < \rho_{\rm c2}$ and
$\rho_{\rm c3} < \rho \le \rho_{\rm c4}$, one finds metastable traffic, i.e. sufficiently small perturbations
will fade away, while large enough ones will grow and cause a breakdown of traffic flow. 
The value of $Q_{\rm out}$ falls into the metastable regime between 
$\rho_{\rm c1}$ and $\rho_{\rm c2}$.
\par
Let us now investigate a bottleneck situation due to inflows
\begin{equation}
\nu_+ = \frac{Q_{\rm rmp}}{nL}
\end{equation} 
along a ramp, where $L$ is the used length of the on-ramp and $n$ the
number of freeway lanes. The corresponding bottleneck strength is, then, 
$\Delta Q = Q_{\rm rmp}/n$. Moreover, let $Q_{\rm up}$ denote 
the traffic flow upstream of the bottleneck and 
\begin{equation}
Q_{\rm tot} = Q_{\rm up} + \Delta Q 
\end{equation}
the total capacity  required downstream of it. 
Then, we expect to always observe free traffic (FT) below the threshold $Q(\rho_{\rm c1})$ of
metastable traffic, i.e. for
\begin{equation}
 Q_{\rm tot} = Q_{\rm up} + \Delta Q < Q(\rho_{\rm c1}) \, .
\end{equation}
In contrast, traffic flow will always be congested, if the maximum flow
$Q_{\rm max} = \max_\rho Q(\rho)$ (the capacity) is exceeded, i.e.
\begin{equation}
 Q_{\rm tot} = Q_{\rm up} + \Delta Q > Q_{\rm max} \, .
\end{equation}
Assuming model parameter for which the maximum flow $Q_{\rm max}$ lies between $\rho_{\rm
  c1}$ and $\rho_{\rm c2}$, traffic states between the two diagonal
lines $Q_{\rm up} + \Delta Q = Q(\rho_{\rm c1})$ and
$Q_{\rm up} + \Delta Q = Q_{\rm max}$ in the $\Delta Q$-over-$Q_{\rm up}$
phase space can be either congested or free, depending
on the initial and boundary conditions. While homogeneous free flow may persist over
long time periods, large perturbations tend to
produce congested states. Extended congested traffic can emerge above the line 
\begin{equation}
Q_{\rm up} =Q_{\rm out} - \Delta Q 
\end{equation}
in the phase diagram, i.e.
if required capacity $Q_{\rm tot}$ is greater than the dynamic capacity $Q_{\rm
  out}$. This line does not have to be parallel to the previously 
mentioned phase boundaries, as $Q_{\rm out}$ may depend 
on the bottleneck strength $\Delta Q$ , but it
depends somewhat on the bottleneck strength $\Delta Q$ (Treiber, Hennecke, and Helbing, 2000;
Helbing, 2001a).  
For $Q(\rho_{\rm c1}) \le Q_{\rm tot} = Q_{\rm up} +
\Delta Q < Q_{\rm out}$, congested traffic states are expected to be 
localized, i.e. they should not extended over long freeway sections. 
\par
The traffic flow $Q_{\rm cong}$ resulting in the congested
area plus the inflow or bottleneck strength $\Delta Q$ are normally
given by the outflow $Q_{\rm out}$, i.e.
\begin{equation}
 Q_{\rm cong} = Q_{\rm out} - \Delta Q
\end{equation}
(if vehicles cannot enter the freeway downstream of the congestion front).
One can distinguish the following cases:
{\em Homogeneous congested traffic} (HCT) can occur,
if the density $\rho_{\rm cong}$ associated with the congested flow 
\begin{equation}
Q_{\rm cong} = Q(\rho_{\rm cong})
\end{equation} 
lies in the stable or meta-stable range 
\begin{equation}
  Q_{\rm cong} \le Q(\rho_{\rm c3}) \, , \quad \mbox{i.e.} \quad 
 \Delta Q \ge Q_{\rm out} - Q(\rho_{\rm c3}) \, .
\end{equation}
Oscillating forms of congested traffic can emerge, if
\begin{equation}
 \Delta Q \le Q_{\rm out} - Q(\rho_{\rm c4}) \quad \mbox{and} \quad
 Q_{\rm up} > Q_{\rm out} - \Delta Q \, .
\end{equation}
That is, lower bottleneck strengths tend to produce
oscillating rather than homogeneous congested flow, and we find
{\em oscillating congested traffic} (OCT), {\em stop-and-go
  waves} (SGW), or {\em moving localized clusters} (MLC). 
In contrast to OCT, stop-and-go
waves are characterized by a sequence of moving jams, between which traffic flows freely.
Under certain conditions, each traffic jam can trigger another one by inducing a small perturbation 
in the inhomogeneous freeway section, 
which propagates downstream as long as it is small, but turns back when it has grown
large enough {\em (``boomerang effect'')}. This requires the downstream traffic flow
to be linearly unstable or at least meta-stable. 
If it is metastable, however, (when the traffic volume $Q_{\rm tot}$
is sufficiently small), i.e. if 
\begin{equation}
 Q(\rho_{\rm c1}) \le  Q_{\rm tot} = Q_{\rm up} + \Delta Q < Q(\rho_{\rm c2}) \, ,
\end{equation}
small perturbations may fade away. In that case, one
either finds a single {\em moving localized cluster} (MLC), or a {\em pinned localized cluster} (PLC)
at the location of the ramp. The latter requires the traffic flow in the upstream section to be stable, i.e.
\begin{equation}
 Q_{\rm up} < Q(\rho_{\rm c1}) \, , 
\end{equation}
so that no traffic jam can survive there. In contrast, moving
localized clusters and triggered stop-and-go waves require 
\begin{equation}
Q_{\rm up} \ge Q(\rho_{\rm c1}) \, .
\end{equation}
\par
In practical situations, things are a little bit more complicated:
\begin{itemize}
\item The phase diagram depicted in Fig.~\ref{Phased} assumes 
a single bottleneck only, while in many cases one is, for example, confronted with a combination of
off- and on-ramps. Therefore, if traffic upstream of an on-ramp is congested, drivers may react to
this by leaving the freeway over the off-ramp. This behavior can suppress a growth of congested
traffic beyond the off-ramp. As a consequence, the upstream front of extended congested
traffic may be pinned, when it reaches the location of an off-ramp,
and OCT states may look like OPLC states, while HCT states may sometimes look
like PLC states.
\item More detailed simulation studies show that there are also multistable areas,
where either PLC, OCT, or FT states can emerge,
depending on the respective initial condition. This is particularly the case, if the
outflow $Q_{\rm out}$ from congested traffic does not exactly agree with $Q(\rho_{\rm c2})$
(Lee, Lee and Kim, 1999; Treiber, Hennecke, and Helbing, 2000).
\item For particular conditions, one also finds traffic states reminding of the ``pinch effect'', 
which is characterized by a spatial coexistence of HCT, OCT, and SGW states
(Helbing, 2001a). More specifically,
the HCT state is found upstream of a bottleneck and turns into an OCT state
further upstream (the ``pinch region''). As the small oscillations travel upstream, they merge to form a SGW 
state. The empirically measured relation between the wavelength of the oscillations and the maximum
vehicle velocity (Kerner, 1998a) is convincingly reproduced by simulations (Helbing {\em et al.}, 2003).
\item In principle, 
phase diagrams can also be generated for more complex freeway geometries with several bottlenecks.
These are, however, not two- but multidimensional (with one additional dimension per bottleneck).
For this reason, there will be many more phases, corresponding to the possible combinations of
the six traffic states FT, PLC, MLC, SGW, OCT, and HCT at different bottlenecks and cases where
extended traffic states influence the traffic states at bottlenecks further upstream. 
All these complex states can be simulated on a computer, when the location-dependent capacities of the
freeway plus the boundary flows and ramp flows are known.
\item In cases of long ramps, vehicles may enter the freeway downstream of the congestion front induced
by it. The downstream flow per lane may, therefore, be higher than $Q_{\rm out}$.
\item The large scattering of traffic flows due to variations in the netto time gaps (see Banks, 1999; Treiber 
and Helbing, 1999; Nishinari {\em et al.}, 2003)
makes it difficult to locate the exact point of empirical traffic states in the phase diagram.
Therefore, the empirical phase boundaries are expected to have a fuzzy appearance and to depend on the respective truck fraction. 
\end{itemize}

\begin{figure}[htbp]
\begin{center}
\end{center}
\caption{Phase points of observed traffic states in the phase space spanned by
the effective on-ramp flow $Q_{\rm rmp}/n$ per freeway lane and the upstream freeway flow $Q_{\rm up}$
per lane (F = free traffic, M = moving localized cluster, S = stop-and-go waves,
O = oscillating congested traffic, L = localized traffic [probably not pinned localized clusters, but
oscillating or homogeneous congested traffic that did not pass the off-ramp]). 
The displayed flow values are averages over the 10 minutes immediately
before the breakdown of traffic flow. Solid lines are supposed to be guides to the eyes to support
a comparison with the theoretical phase diagram displayed in Fig.~\ref{Phased}.}
\label{epd}
\end{figure}

Despite of the above complications, the empirical phase diagram depicted in Fig.~\ref{epd}
is surprisingly well compatible with the theoretical one (see Fig.~\ref{Phased}): 
\begin{itemize}
\item For values of the overall flow $Q_{\rm tot} = Q_{\rm up} + Q_{\rm rmp}/(nL)$ 
below 1300 vehicles per hour and lane, traffic is always free, while above 2100 vehicles
per hour and lane, it is always congested. 
\item Moving localized clusters occur for upstream flows above 1300 vehicles per  hour and lane
and for overall flows $Q_{\rm tot}$ below 1800 vehicles per hour and lane.
\item For $1800\mbox{ veh./h/lane} \le Q_{\rm tot} \le 2100$ veh./h/lane and $Q_{\rm up}
\ge 1300$ veh./h/lane, we find a coexistence of free traffic, oscillating congested traffic,
and stop-and-go waves. 
\item Homegeneous congested traffic is only found after serious accidents with lane closures,
with a bottleneck strength $\Delta Q \ge 900$ vehicles per hour and lane (not shown).  
\end{itemize}
Altogether, there is a good agreement between the theoretical and empirical traffic states,
and  the empirical phase diagram is qualitatively compatible with
the theoretical phase diagram. The only debatable point is the interpretation of the
localized states (L) as extended congested traffic states that did not pass the off-ramp
due to drivers leaving the freeway. We will, therefore, discuss another, flow-conserving bottleneck
caused by a gradient (see Fig.~\ref{grad}). In this case, stationary localized clusters must be actually 
pinned localized clusters (PLC), but the bottleneck strength $\Delta Q$ is hard to measure. For
this reason, we do not have a two-dimensional phase diagram, but an intersection through it.
Assuming that the bottleneck strength can be estimated by $\Delta Q \approx C_1
Q_{\rm up}$,  the phase diagram predicts the following sequence of traffic states with 
growing freeway flows $Q_{\rm up}$: 
free traffic $\longrightarrow$ pinned localized clusters
or free traffic $\longrightarrow$ moving localized clusters or free traffic $\longrightarrow$
oscillating congested traffic, stop-and-go waves, moving localized clusters, or free traffic
$\longrightarrow$ oscillating congested traffic or stop-and-go waves $\longrightarrow$
oscillating congested traffic or homogeneous congested traffic $\longrightarrow$ homogeneous
congested traffic. Despite the wide scattering of traffic flow data, this theoretically predicted
sequence is compatible with our empirical observations (see Fig.~\ref{Phased}). The empirical
data suggest the following values for the gradient bottleneck: $Q(\rho_{\rm c1}) \approx 1700$~veh./h/lane,
$Q_{\rm out} \approx 1900$~veh./h/lane, and $C_1 \approx 0.06$.
\begin{figure}[htbp]
\begin{center}
\end{center}
\caption{
Empirical traffic states at  the gradient
around kilometer 479 on the German freeway A5 in direction North. Despite the considerable
scattering, one observes a tendency that pinned localized clusters (P) occur at higher
upstream flows $Q_{\rm up}$ than free traffic (F), and even higher flows tend to produce
moving localized clusters (M), stop-and-go waves (S), or oscillating congested traffic (O).}
\label{grad}
\end{figure}

\subsection{Relevance for Other Traffic Models}

Is the non-local, gas-kinetic-based traffic model (GKT model) the 
only one that produces qualitatively correct traffic states
and phase diagrams? The answer is no. According to the above theoretical explanation of the
phase boundaries, the same properties are expected for all traffic models with a similar
instability diagram. It is particularly essential to have a density regime $\rho_{\rm c1} \le \rho < \rho_{\rm c2}$,
in which traffic is metastable. Moreover, the outflow $Q_{\rm out}$ from congested traffic needs to be
smaller than the maximum homogeneous traffic flow. Finally, there should be a critical density $\rho_{\rm c3}$, 
above which traffic is metastable or stable. The GKT model has these features in common with, for example,
the Kerner-Konh\"auser model (1994), the optimal velocity model (Bando {\em et al.}, 1995b), and
the intelligent driver model (Treiber {\em et al.}, 2000). It is, therefore, not surprising
that qualitatively the same traffic states and phase diagrams have been found for them. 
The reader can check this by means of the Java simulation applet of an on-ramp scenario 
supplied at {\tt http://www.mtreiber.de}. By variation of the main flow and the on-ramp
flow, it is possible to produce different kinds of traffic states. 
\par
This suggests that phase diagrams of traffic states may be used to classify traffic models, no matter whether
they are microscopic, cellular automata, car-following, gas-kinetic or macroscopic models
(see Sec.~\ref{Prev}). According to the above,
the IDM model, which is a microscopic car-following model, may be called equivalent to the macroscopic 
Kerner-Konh\"auser model or the GKT model, when it comes to qualitative features of traffic states. 
Other models have different instability and phase diagrams: 
\begin{itemize}
\item Models without a metastable density regime $\rho_{\rm c1} \le \rho < \rho_{\rm c2}$ 
will not show the localized MLC and PLC states, although they may still possess SGW states. 
According to Krau{\ss} (1998), traffic models show the observed
metastable traffic and a capacity drop only, if the typical maximal acceleration 
is not too large and if the deceleration strength is moderate. 
\item Models without a stable high-density range 
will not show a HCT state.
For example, if the model displays stable traffic at small densities and unstable
traffic at high densities, one expects free traffic (FT) and 
oscillating traffic (OCT or SGW). This was observed by Emmerich and Rank (1995)
(see also Diedrich \etal, 2000; Cheybani \etal, 2001). 
\item Models without a linearly unstable regime
(such as the Burgers equation or the LWR model) will not produce 
emergent oscillating states (OCT or SGW). For the totally asymmetric
exclusion process (TASEP), a simple microscopic particle hopping model,
this has been shown to be in agreement with numerical
(Janowsky and Lebowitz, 1992) and exact analytical results (Sch\"utz, 1993).
\end{itemize}
Although these model classes have different phase diagrams, all of them produce extended congested traffic
states (vehicle queues) at bottlenecks. 

\section{Summary and Discussion} \label{Discussion}

In this contribution, we have presented results of a systematic analysis of empirical traffic states
of a 30 kilometer long section of the German freeway A5 near Frankfurt. Despite two
freeway intersections, there are two approximately 10 kilometer long
freeway sections without any on- or off-ramps. Along the freeway section, one finds a rich variety of 
congested traffic states, but the great majority of them can be interpreted as a spatial coexistence of
altogether six different traffic states: free traffic (FT), pinned localized clusters (PLC), moving
localized clusters (MLC), stop-and-go waves (SGW), oscillating congested traffic (OCT), 
and homogeneous congested traffic (HCT). Note that the existence of homogeneous congested
traffic points to stable traffic flow at very high densities or very low velocities.
\par
The typical traffic pattern depends on the
overall flow (i.e. the upstream freeway flow plus the ramp flow) and, therefore, on the day of the week,
but due to multi-stability, initial and boundary conditions are relevant as well. 
The most frequent states at the investigated freeway section
are PLC and OCT states, while HCT occurs mainly after serious
accidents with lane closures or during public holidays.  The downstream fronts
of these congestion patterns are located at bottlenecks. 
\par
Note that bottlenecks may have different
origins: on-ramps, reductions in the number of lanes, accidents
(even in opposite lanes because of rubbernecks), speed limits, road works,
gradients, curves, bad road conditions (possibly due to rain, fog, or ice),
bad visibility (e.g., because of blinding sun or tunnel entrances), diverges 
(``negative'' perturbations, see Helbing {\em et al.}, 2003), 
weaving flows by vehicles trying to switch to the slow
exit lane, or congestion on the off-ramp, see  Daganzo \etal, 1999;
Lawson \etal, 1999; Mu\~noz and Daganzo, 1999; Daganzo, 2002a).
Moving bottlenecks due to 
slow vehicles are possible as well (Gazis and Herman, 1992), 
leading to a forward movement of the respective downstream 
congestion front (see Fig.~\ref{phantom}). Finally, in cases of {\em two} subsequent inhomogeneities 
of the road, there are forms of congested traffic in which both,
the upstream and downstream fronts are locally 
fixed  (Treiber \etal, 2000; Lee \etal, 2000; Kerner, 2000a). 
\par
For the identification of the traffic states, we have used an
adaptive smoothing method which interpolates and smoothes traffic data from successive
freeway sections, taking into account the propagation speeds of perturbations in free and congested
traffic. The method is based on measurements of average vehicle speeds by double-loop detectors,
which are quite reliable. Our inverse speed representation reminds of density plots, which 
is particularly intuitive. It also allows one to identify small structures such as the boomerang
effect, which is one of the mechanisms of congestion formation: 
Perturbations in free traffic travel downstream as long as they are small. However, in some cases
they grow in amplitude and change their speed and direction of motion, similar to a boomerang.
When the considerably amplified perturbation reaches a bottleneck, it can cause a breakdown
of traffic flow, as the outflow from congested traffic is smaller than the maximum homogeneous
flow. 
\par
In contrast to Kerner's empirical analysis of the same freeway stretch, our study is based
on an analysis of the spatio-temporal traffic dynamics rather than the dynamics at single freeway
sections or comparisons of the dynamics at subsequent sections. The growth of small
perturbations in the absence of ramps demonstrated by us is not compatible with
the Lighthill-Whitham-Richard model, while several second-order models can reproduce them. 
Growing perturbations may be also described by first-order models
with an additional diffusion term $D(\rho) \, \partial^2 \rho/\partial x^2$, which becomes
negative in an unstable range at medium densities: 
\begin{equation}
 \frac{\partial \rho(x,t)}{\partial t} + \frac{\partial Q(\rho(x,t))}{\partial x} = D(\rho(x,t)) 
 \frac{\partial^2 \rho(x,t)}{\partial x^2}
 + \mbox{higher order terms} + \nu_+(x,t) - \nu_-(x,t) \, .
\end{equation}
Such models have common features with second-order models and
can be theoretically supported (Nelson, 2000; Helbing, 2001a), but the stability of their solutions
as a function of higher-order terms
requires further investigations. In any case, the congested traffic states
identified by us are in good agreement with predictions of some second-order macroscopic traffic models
and some microscopic car-following models. Moreover, the dependence of the resulting traffic
patterns on the respective freeway and ramp flows was qualitatively correct despite the
variation of the flow data.
\par
The erratic scattering of flow-density data is partly an effect of aggregation by
loop detector data (i.e. the measurement procedure, see Helbing {\em et al.}, 2003;
Treiber and Helbing, 2003). However, the main
reason seems to be the broad distribution of the individual netto time gaps 
(Banks, 1999; Neubert {\em et al.}, 1999; Treiber and Helbing, 1999; Tilch and Helbing, 2000; Helbing {\em et al.},
2002, 2003; Nishinari {\em et al.}, 2003). The variation of the density $\rho$
{\em and} the average netto time gap $T$
together can explain changes in the flow with a correlation of more the 90\% (Nishinari {\em et al.}, 2003):
According to Eq.~(\ref{Qt}), a variation of $T$ causes
a variation in the slope of the jam line $J(\rho)$, while variations of $\rho$ cause variations 
along the respective jam line.
Altogether this can explain variations perpendicular and parallel to the average jam line.
A simulation of a mixture of cars and trucks had already pointed in
this direction (Treiber and Helbing, 1999). In summary, an erratic and wide scattering of flow-density data
can be reproduced, taking into account the heterogeneity in the driver-vehicle behavior
due to the wide distribution of time gaps. Therefore, it does not contradict models with a 
fundamental diagram.

\subsection*{Acknowledgments}

The authors would like to thank for financial support by the Volkswagen AG
within the BMBF research initiative INVENT, 
the {\it Hessisches Landesamt f\"ur Stra{\ss}en und Verkehrswesen}
for providing {the} freeway data, and Martin Treiber for valuable discussions.

\end{document}